\documentclass[reprint,amsmath,amssymb,aps,superscriptaddress]{revtex4-2}
\pdfoutput=1

\usepackage{tikz}
\usepackage{graphicx}
\usepackage{dcolumn}
\usepackage{bm}
\usepackage{braket}
\usepackage[lined,boxruled]{algorithm2e}
\usepackage{multirow}

\usepackage{qcircuit}
\usepackage{hyperref}
\usepackage[capitalise]{cleveref}
\usepackage{siunitx}

\newcommand{\fswap}{\mathcal{F}}
\newcommand{\xx}{X_{\pi/2}}
\newcommand{\iswap}{i\text{SWAP}}
\newcommand{\swap}{\text{SWAP}}

\newcommand{\cphase}[1]{\text{CPHASE}(#1)}

\def\cz/{{\ensuremath{{CZ}}}}
\def\cs/{{\ensuremath{{CS}}}}
\def\csd/{{\ensuremath{{CS}^\dagger}}}
\def\ct/{{\ensuremath{{CT}}}}

\begin{document}

\preprint{APS/123-QED}

\title{Optimized fermionic SWAP networks with equivalent circuit averaging for QAOA}

\author{Akel Hashim}
    \thanks{These authors contributed equally to this work.}
    \affiliation{Quantum Nanoelectronics Laboratory, Department of Physics, University of California at Berkeley, Berkeley, CA 94720, USA}
    \affiliation{Graduate Group in Applied Science and Technology, University of California at Berkeley, Berkeley, CA 94720, USA}
    \affiliation{Computational Research Division, Lawrence Berkeley National Lab, Berkeley, CA 94720, USA}
\author{Rich Rines}
    \thanks{These authors contributed equally to this work.}
    \affiliation{Super.tech, Chicago, IL 60615, USA}
\author{Victory Omole}
    \affiliation{Super.tech, Chicago, IL 60615, USA}
\author{Ravi K. Naik}
    \affiliation{Quantum Nanoelectronics Laboratory, Department of Physics, University of California at Berkeley, Berkeley, CA 94720, USA}
    \affiliation{Computational Research Division, Lawrence Berkeley National Lab, Berkeley, CA 94720, USA}
\author{John Mark Kreikebaum}
    \thanks{Now at Google Quantum AI, Mountain View, CA, USA.}
    \affiliation{Quantum Nanoelectronics Laboratory, Department of Physics, University of California at Berkeley, Berkeley, CA 94720, USA}
    \affiliation{Materials Sciences Division, Lawrence Berkeley National Lab, Berkeley, CA 94720, USA}
\author{David I. Santiago}
    \affiliation{Computational Research Division, Lawrence Berkeley National Lab, Berkeley, CA 94720, USA}
\author{Frederic T. Chong}
    \affiliation{Super.tech, Chicago, IL 60615, USA}
    \affiliation{University of Chicago, Chicago, IL 60637, USA}
\author{Irfan Siddiqi}
    \affiliation{Quantum Nanoelectronics Laboratory, Department of Physics, University of California at Berkeley, Berkeley, CA 94720, USA}
    \affiliation{Computational Research Division, Lawrence Berkeley National Lab, Berkeley, CA 94720, USA}
    \affiliation{Materials Sciences Division, Lawrence Berkeley National Lab, Berkeley, CA 94720, USA}
\author{Pranav Gokhale}
    \thanks{Email correspondence: pranav@super.tech}
    \affiliation{Super.tech, Chicago, IL 60615, USA}

\date{\today}

\begin{abstract}
The fermionic SWAP network is a qubit routing sequence that can be used to efficiently execute the Quantum Approximate Optimization Algorithm (QAOA). Even with a minimally-connected topology on an $n$-qubit processor, this routing sequence enables $\mathcal{O}(n^2)$ operations to execute in $\mathcal{O}(n)$ steps. In this work, we optimize the execution of fermionic SWAP networks for QAOA through two techniques. First, we take advantage of an overcomplete set of native hardware operations [including 150 ns controlled-$\frac{\pi}{2}$ phase gates with up to 99.67(1)\% fidelity] in order to decompose the relevant quantum gates and SWAP networks in a manner which minimizes circuit depth and maximizes gate cancellation. Second, we introduce Equivalent Circuit Averaging, which randomizes over degrees of freedom in the quantum circuit compilation to reduce the impact of systematic coherent errors. Our techniques are experimentally validated on the Advanced Quantum Testbed through the execution of QAOA circuits for finding the ground state of two- and four-node Sherrington–Kirkpatrick spin-glass models with various randomly sampled parameters. We observe a $\sim\!60$\% average reduction in error (total variation distance) for QAOA of depth $p = 1$ on four transmon qubits on a superconducting quantum processor.
\end{abstract}

\maketitle

\section{Introduction}
A key challenge for scaling near-term quantum computers to address practical problems is limited qubit connectivity. While qubit mapping techniques can mitigate this limitation, recent results suggest that any mismatch between hardware connectivity and connectivity required for specific applications can erase the potential for a quantum speedup \cite{franca2020limitations, wang2020noise}. This poses a particular challenge for superconducting quantum hardware which --- despite the advantages of fast operation speed, high gate fidelity, and scalable fabrication --- generally has the disadvantage of sparse nearest-neighbor qubit connectivity.

The fermionic SWAP network, introduced in Ref.~\cite{kivlichan2018quantum} and studied further in Refs.~\cite{o2019generalized, hagge2020optimal}, offers a promising path forward for coping with limited connectivity. In fact, the fermionic SWAP network requires only minimal linear connectivity between qubits; any additional qubit couplings are unnecessary. This property is well-suited to superconducting qubits where it has the additional advantage of minimizing the effect of crosstalk due to frequency crowding \cite{ding2020systematic}.

Qubit routing in an $n$-qubit fermionic SWAP network follows a sequence of $n - 1$ steps, incurring $\mathcal{O}(n)$ total quantum circuit depth. This linear cost suffices to carry out all $\mathcal{O}(n^2)$ pairwise interactions between qubits, even for linearly-arranged qubits. By contrast, naive qubit routing approaches would require $\mathcal{O}(n^3)$ circuit depth to perform all of the necessary operations, because each of the $\mathcal{O}(n^2)$ pair-wise interactions would be serialized and would incur an $\mathcal{O}(n)$ SWAP overhead. The quadratic advantage in circuit depth offered by fermionic SWAP networks persists even in comparison to state-of-the-art qubit routing \cite{tomesh2020coreset}.

There are numerous applications of fermionic SWAP networks, broadly corresponding to evolution under a fully-connected Hamiltonian comprising mutually commuting terms. Examples include the Sherrington-Kirkpatrick spin-glass model \cite{farhi2019quantum}, Max-Cut for use cases like VLSI circuit design \cite{barahona1988application}, and k-means clustering on large datasets with coresets \cite{tomesh2020coreset}. Furthermore, Hamiltonian evolution is at the heart of many noisy intermediate-scale quantum (NISQ) \cite{preskill2018quantum} algorithms such as the Quantum Approximate Optimization Algorithm (QAOA) \cite{farhi2014quantum} and its derivatives \cite{bravyi2020obstacles, hadfield2019quantum, wurtz2021counterdiabaticity}, making the implementation of fermionic SWAP networks invaluable for near-term applications. In addition, fermionic SWAP networks will be favorable for noise mitigation approaches involving virtual distillation \cite{koczor2020exponential}, in which multiple copies of a quantum state can be arranged in parallel registers with linear connectivity \cite{huggins2020virtual}.

Given the fundamental importance of fermionic SWAP networks to many quantum applications, it is important to fully optimize their execution. Here, we introduce and apply two compilation techniques that improve their performance. The first technique employs a richer gateset than enabled by standard QASM (Quantum Assembly) representation for circuit decomposition. 
The second technique, which we term Equivalent Circuit Averaging (ECA), involves randomizing circuit decomposition over degrees of freedom in compilation to mitigate the impact of systematic coherent errors. Both of these techniques are validated at the Advanced Quantum Testbed at Lawrence Berkeley National Laboratory.

The rest of this paper is organized as follows. Section~\ref{sec:aqt} describes the Advanced Quantum Testbed's hardware. Section~\ref{sec:optimized_fermionic_swap} presents our optimized gate decompositions for the Hadamard, SWAP, and Fermionic SWAP operations. Section~\ref{sec:cycle_benchmarking} presents results from cycle benchmarking of our optimized gate sequences. Section~\ref{sec:qaoa} examines the application of fermionic SWAP networks to QAOA, and Section~\ref{sec:equivalent_circuit_averaging} introduces Equivalent Circuit Averaging for this application. Section~\ref{sec:conclusion} concludes. Appendices \ref{appendix_a}~and~\ref{appendix_b} detail single-qubit and two-qubit parameters for the Advanced Quantum Testbed. Finally, Appendix~\ref{appendix_c} presents examples of the full fermionic SWAP network circuits that we executed.

\section{The Advanced Quantum Testbed} \label{sec:aqt}
The experiments in this work were performed on four fixed-frequency transmon \cite{koch2007charge} qubits (labeled Q4, Q5, Q6, and Q7; see Table \ref{tab:table_sqp} in Appendix \ref{appendix_a}) on an eight-qubit superconducting quantum processor ($\texttt{AQT@LBNL Trailblazer8-v5.c2}$) at the Advanced Quantum Testbed \cite{AQT} (AQT). The qubits are coupled to nearest-neighbors via fixed-frequency resonators in a ring-geometry.

Arbitrary single-qubit $SU(2)$ gates are typically implemented using physical $X_{\pi/2}$ gates (via resonant Rabi-driven pulses) and virtual $Z_{\theta}$ gates (via phase shifts between physical pulses) \cite{mckay2017efficient}:
\begin{equation}
    \label{eq:zxzxz}
    U(\alpha, \beta, \gamma) = Z_{\alpha - \pi/2}X_{\pi/2}Z_{\pi - \beta}X_{\pi/2}Z_{\gamma - \pi/2}.
\end{equation}
This $ZXZXZ$-decomposition reduces the time and complexity involved in calibrating and benchmarking single-qubit gates. The disadvantage is that every computational single-qubit gate is actually composed of two physical $X_{\pi/2}$ pulses, each 30 ns in duration; thus, every single-qubit gate (cycle) in a circuit takes 60 ns by default, even if the gate could be implemented with only a single $X_{\pi/2}$ pulse. This needlessly increases circuit depth, leaving the qubits more susceptible to decoherence.

Two-qubit entangling operations are achieved using a tunable ZZ-coupling via off-resonant drives \cite{mitchell2021hardware, wei2021quantum} between neighboring qubits, which is used to implement controlled-$Z$ ($CZ$) operations between all qubit pairs, as well as controlled-$S$ ($CS$) and controlled-$S^\dagger$ ($CS^\dagger$) gates. The duration of our two-qubit $CZ$ gate is 200 ns, which is limited by the drive-induced decoherence discussed in Ref.~\cite{mitchell2021hardware}. However, because the $CS$ or $CS^\dagger$ gate performs half the rotation of a $CZ$, it can be implemented in less time than the $CZ$. We calibrate and measure a process infidelity of 4.3(1)$\times 10^{-3}$ for a 150 ns $CS$ gate between qubits (Q5, Q6), and process infidelities of 5.0(1)$\times 10^{-3}$ and 3.3(1)$\times 10^{-3}$ for a 150 ns $CS^\dagger$ gate between qubits (Q4, Q5) and (Q6, Q7), respectively (see Table \ref{tab:table_tqp} in Appendix \ref{appendix_b}), which is $\sim\!100$ ns faster with an error rate that is $\sim\!2\times$ lower than previously measured for superconducting qubits \cite{garion2021experimental}.

\section{Optimized Gate Decompositions}\label{sec:optimized_fermionic_swap}
\subsection{Optimized Hadamard and SWAP}\label{sec:optimized-hadamard-and-swap}
We first optimize decompositions for the Hadamard ($H$) and SWAP operations. The $H$ gate has two equivalent decompositions using the $\{X_{\pi/2}, Z_{\theta}\}$ basis:
\begin{equation} \footnotesize
\label{eq:circuit:h_std}
    {\Qcircuit @C=.4em @R=.2em {
      & \gate{H} & \qw && \text{\!=\!} &&& \gate{X_{\pi/2}} & \gate{Z_{\pi/2}} & \gate{X_{\pi/2}} & \qw \\
    }}_,
\end{equation}
\begin{equation} \footnotesize
\label{eq:circuit:h_opt}
    {\Qcircuit @C=.4em @R=.2em {
      & \gate{H} & \qw && \text{\!=\!} &&&  \gate{Z_{\pi/2}} & \gate{X_{\pi/2}} & \gate{Z_{\pi/2}} & \qw \\
    }}_.
\end{equation}
The standard $ZXZXZ$-decomposition of the $H$ gate corresponds to Eq.~\ref{eq:circuit:h_std}. While this is a valid decomposition,  Eq.~\ref{eq:circuit:h_opt} is preferable because it requires a single physical $X_{\pi/2}$ pulse instead of two. Therefore, the optimized Hadamard halves the duration of the gate, taking only 30 ns instead of 60 ns.

Next, we consider the SWAP operation. The default decomposition of the SWAP  is
\begin{equation}
\label{eq:circuit:vanilla_swap}
{\Qcircuit @C=.5em @R=.2em {
  & \qswap\qwx[1] & \qw &&&&& \qw & \ctrl{2} & \targ     & \ctrl{2} & \qw &&&&& \qw & \qw & \ctrl{2} & \gate{H} & \ctrl{2} & \gate{H} & \ctrl{2} & \qw & \qw \\
  &&&& \text{\Large=}   &&&&&&&&& \text{\Large=}  \\
  & \qswap\qwx[-1] & \qw &&&&& \qw & \targ   & \ctrl{-2} & \targ & \qw    &&&&& \qw & \gate{H} & \ctrl{0} & \gate{H}  & \ctrl{0} & \gate{H} & \ctrl{0} & \gate{H} & \qw  \\
}}_,
\end{equation}
where $\raisebox{.6em}{\Qcircuit @C=0em @R=.4em { \; & \ctrl{1} \\ \; & \ctrl{0} \\ }}$ indicates the \cz/ gate. The middle circuit represents the standard QASM decomposition of a SWAP, which involves three alternating $CX$ gates. No further decompositions are possible until we go below the level of QASM \cite{gokhale2021faster}. Each $CX$ decomposes to AQT's two-qubit $CZ$ basis gate by invocation of the identity $CX(q_c, q_t) = H(q_t) CZ(q_c, q_t) H(q_t)$, where $q_c$ ($q_t$) is the control (target) qubit. We immediately see that applying the optimized $H$ leads to an improvement: the total SWAP duration is reduced by $4 \times 30\;\text{ns} = 120\;\text{ns}$, and the number of required $X_{\pi/2}$ physical pulses is halved from 12 to 6.

We can optimize even further by applying a transposition identity to move the bottom-right $H$ to the top-left. This identity reduces the total SWAP duration by an additional 30 ns, since the two ``edge'' $H$ gates become parallelized. After annihilating all virtual rotations arising from \cref{eq:circuit:h_opt} via commutation identities, we have the final optimized SWAP:
\begin{equation}
\label{eq:circuit:swap}
{\Qcircuit @C=.5em @R=.2em {
  & \qswap\qwx[1] & \qw &&&&& \qw
  & \gate{\xx} & \ctrl{2} & \gate{\xx} & \ctrl{2} & \gate{\xx} & \ctrl{2} & \qw \\
  &&&& \text{\Large=} \\
  & \qswap\qwx[-1] & \qw &&&&& \qw 
  & \gate{\xx} & \ctrl{0} & \gate{\xx} & \ctrl{0} & \gate{\xx} & \ctrl{0} & \qw \\
}}_.
\end{equation}
We deployed these optimized Hadamard and SWAP decompositions through the SuperstaQ platform \cite{superstaq}, which can target AQT hardware. Section~\ref{sec:cycle_benchmarking} presents cycle benchmarking \cite{erhard2019characterizing} results for these optimizations.

\subsection{Background on Fermionic SWAP}
The core operation needed in a fermionic SWAP network is the fermionic SWAP gate, defined as the unitary operation below, with input parameter $\theta$:
\begin{equation} \label{eq:f_swap_gate}
  \fswap_\theta =
  \begin{pmatrix}
  1 & 0 & 0 & 0 \\
  0 & 0 & e^{i\theta} & 0 \\
  0 & e^{i\theta} & 0 & 0 \\
  0 & 0 & 0 & 1
  \end{pmatrix}.
\end{equation}
The standard QASM-decomposed quantum circuit implementation of $\fswap_\theta$ comprises three $CX$ gates and a single-qubit $Z_{\theta}$ rotation \cite{tomesh2020coreset}:
\begin{equation}
{\Qcircuit @C=.8em @R=.0em {
  & \multigate{2}{\mathcal{F}_\theta} & \qw & & & \qw
  & \ctrl{2} & \targ & \qw & \ctrl{2} & \qw \\
  & \nghost{\mathcal{F}_\theta} &     & \text{\Large=} & \\
  & \ghost{\mathcal{F}_\theta}        & \qw & & & \qw
  & \targ & \ctrl{-2} & \gate{Z_\theta} & \targ & \qw \\
}}_.
\end{equation}
It is possible to boost performance beyond this decomposition by leveraging knowledge of the target hardware's underlying native gate set. For example, \cite{harrigan2021quantum} compiled the fermionic SWAP operation directly down to 3 native two-qubit Sycamore ($SYC$) gates, rather than recompiling each $CX$ down to $SYC$ gates. Relatedly, \cite{abrams2019implementation} developed a parametric implementation of the fermionic SWAP via access to a native $XY(\theta)$ gate (with duration independent of $\theta$) and a native $CZ$ gate.

However, in these examples, the total duration of the fermionic SWAP operation is always constant, regardless of $\theta$. This leaves room for improvement. For example, it has been shown that access to a parametric $CZ$ [i.e.~$\cphase{\phi}$] yields significant improvements for the decomposition of many quantum operations \cite{barron2020microwave}; in the next subsection we demonstrate that this is true for the fermionic SWAP operation as well. However, there are experimental obstacles to tuning a high-fidelity parametric gate with variable duration. For example, \cite{gokhale2020optimized} noted ramp effects at small $\theta$ for a parametric cross-resonance gate.

Rather than incurring the calibration overhead of a parametric gate with variable duration like  $\cphase{\phi}$, we instead focus on the optimization opportunities from an \textit{overcomplete} discrete two-qubit gate set. Concretely, we next examine the optimized $\fswap_\theta$ decompositions possible when we have access to both a $CZ$ and $CS = \sqrt{CZ}$ gate, where the $CS$ is faster than the $CZ$ gate.

\begin{figure}[h] 
    \begin{equation*}
    {\Qcircuit @C=.4em @R=0.2em {
      & \qw & \multigate{6}{e^{i\gamma H}} &\qw& \ghost{\text{\Large=}} 
      & \qw & \multigate{2}{\fswap_{\theta_{01}}}
      & \qw & \multigate{2}{\fswap_{\theta_{13}}} & \qw & \qw & \qw \\
      & \\
      & \qw & \ghost{e^{i\gamma H}} &\qw& \ghost{\text{\Large=}} 
      & \qw & \ghost{\fswap_{\theta_{01}}} & \multigate{2}{\fswap_{\theta_{03}}}
      &       \ghost{\fswap_{\theta_{13}}} & \multigate{2}{\fswap_{\theta_{12}}} & \qw & \qw \\
      &&&& \text{\Large=} \\
      & \qw & \ghost{e^{i\gamma H}} &\qw& \ghost{\text{\Large=}} 
      & \qw & \multigate{2}{\fswap_{\theta_{23}}} & \ghost{\fswap_{\theta_{03}}} 
      &       \multigate{2}{\fswap_{\theta_{02}}} & \ghost{\fswap_{\theta_{12}}} & \qw & \qw \\
      & \\
      & \qw & \ghost{e^{i\gamma H}} &\qw& \ghost{\text{\Large=}} 
      & \qw & \ghost{\fswap_{\theta_{23}}}
      & \qw & \ghost{\fswap_{\theta_{02}}} & \qw & \qw & \qw \\
    }}
    \end{equation*}
\caption{Fermionic SWAP network implementing the Hamiltonian evolution $e^{i\gamma H}$ for a four-node Sherrington-Kirkpatrick model $H=\sum_{i<j<4}J_{ij}Z_iZ_j$, where $\theta_{ij}=\gamma J_{ij}$. Note that the qubit order is reversed after the operation}
\label{fig:fswap:network}
\end{figure}
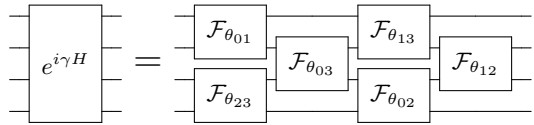

In typical applications, multiple fermionic SWAP gates are arranged into a nearest-neighbor fermionic SWAP \emph{network} that carries out $t = 0, \dots, n - 1$ steps. Each step alternates between an odd and even pattern. At steps with odd $t$, each neighboring qubit pair with indices $(2k, 2k + 1)$ for $k \in [0, \frac{n}{2}]$ is entangled in accordance with a target Hamiltonian and then SWAPped. Even-$t$ steps perform this interaction for qubit pairs with indices $(2k + 1, 2k + 2)$. Note that each step is highly parallel, with $\sim n/2$ operations occurring simultaneously.   A prototypical example is shown in Fig.~\ref{fig:fswap:network}, which implements the Hamiltonian evolution $e^{i\gamma H}$ corresponding to a Sherrington-Kirkpatrick spin-glass model $H=\sum_{i<j<n}J_{ij}Z_iZ_j$ on $n=4$ nodes. The utility of a fermionic SWAP network is that it efficiently generates an all-to-all interaction with a linear-depth circuit of nearest-neighbor interactions, where each interaction is a single fermionic SWAP gate corresponding to one of the commuting weight-2 terms in $H$.

\subsection{Optimized Fermionic SWAP Gates}\label{sec:optimized_fermionic_swap_gates}

We now introduce optimized, $\theta$-dependent decompositions of the fermionic SWAP gate $\fswap_\theta$, taking advantage of an \textit{overcomplete} two-qubit gate set consisting of both $CZ$ and $CS$ or $CS^\dagger$ gates.

For $\theta\in\{0,\pi\}$, the fermionic SWAP unitary in Eq.~\ref{eq:f_swap_gate} is equivalent (up to virtual phases) to the standard SWAP gate, and so can be decomposed using the optimized SWAP described in \cref{sec:optimized-hadamard-and-swap}. In the general case, using the optimized Hadamard (c.f.~Eq.~\ref{eq:circuit:h_opt}) and virtual $Z$ rotations, the baseline fermionic SWAP decomposition still requires three \cz/ and six $\xx$ gates:
\begin{equation}
\label{eq:fswap:cz}
{\Qcircuit @C=.38em @R=.1em {
  & \multigate{2}{\!\fswap_\theta\!} & \qw &&&&
  & \qw        & \ctrl{2} & \gate{\xx} & \ctrl{2} & \qw             & \gate{\xx} & \ctrl{2} & \qw & \qw \\
  &  &&& \text{\Large=} \\
  & \ghost{\!\fswap_\theta\!} & \qw &&&&
  & \gate{\xx} & \ctrl{0} & \gate{\xx} & \ctrl{0} & \gate{Z_{-\theta}} & \gate{\xx} & \ctrl{0} & \gate{\xx} & \qw \\
}}_.
\end{equation}
Unfortunately, in the general case the first and final $\xx$ gates in \cref{eq:fswap:cz} cannot be parallelized as they are in the optimized SWAP, and so both contribute to the depth of the standalone fermionic SWAP circuit. In \cref{sec:fswap:schedule}, we will show that this overhead can be mitigated in the context of a full fermionic SWAP network.

A second special case exists for $\theta=\pm\pi/2$, in which the fermionic SWAP is equivalent to the $\iswap$ ($\iswap^\dagger$) gate and requires just two \cz/ gates. Again employing the optimized Hadamard, it can be decomposed:
\begin{equation}
\label{eq:fswap:iswap}
{\Qcircuit @C=.5em @R=.2em {
  & \qw & \qswap\qwx[1] & \qw & \qw &&&&& \qw
  & \gate{\xx} & \ctrl{2} & \gate{\xx} & \ctrl{2} & \gate{\xx} & \gate{Z_{\pm\pi/2}} & \qw \\
  & {\pm i\;\;} && &&& \text{\Large=} \\
  & \qw & \qswap\qwx[-1] & \qw & \qw &&&&& \qw 
  & \gate{\xx} & \ctrl{0} & \gate{\xx} & \ctrl{0} & \gate{\xx} & \gate{Z_{\pm\pi/2}} & \qw \\
}}_.
\end{equation}

If we have access to a parameterized $\cphase\phi$ gate, we can naturally generalize the optimized SWAP and $\iswap$ decompositions to all fermionic SWAP circuits. Setting $\phi=\pi-2\theta$,
\begin{equation}
\label{eq:fswap:crz}
{\Qcircuit @C=.4em @R=.1em {
  & \multigate{2}{\!\fswap_\theta\!} & \qw &&&&
  & \gate{\xx} & \ctrl{2} & \gate{\xx} & \ctrl{2} & \gate{\xx} & \ctrl{2} & \gate{Z_\theta} & \qw \\
  &  &&& \text{\Large=} \\
  & \ghost{\!\fswap_\theta\!} & \qw &&&&
  & \gate{\xx} & \ctrl{0} & \gate{\xx} & \ctrl{0} & \gate{\xx} & \gate{Z_{\pi-2\theta}} & \gate{Z_\theta} & \qw \\
}}_,
\end{equation}
correctly generates the three-\cz/ optimized SWAP for $\theta\in\{0,\pi\}$ and the two-\cz/ optimized $\iswap$ for $\theta=\pm\pi/2$. Assuming the gate time of any $\cphase\phi$ gate to be proportional to $\phi$ for intermediary $0\le\phi\le\pi$, the total time for $\fswap_\theta$ is the same as
$2+2|\theta\bmod\pi - \pi/2|/\pi$ \cz/ gates.

\begin{figure}
  \begin{center}
    \vspace{10pt}
    \begin{tikzpicture}[scale=2.0,cap=round,>=latex]
      \tikzstyle{color1}=[draw=red];
      \tikzstyle{color2}=[draw=blue!80];
      \draw[->] (-1.3cm,0cm) -- (1.3cm,0cm);
      \draw[->] (0cm,-1.3cm) -- (0cm,1.3cm);
      \draw[gray] (0cm,0cm) circle(1cm);

      \draw[thick,color1] (0.6cm,0.85cm) -- (0.7cm,0.95cm)
        node[inner sep=2pt,anchor=south west,fill=white] {\color{red}$\bm{2\times \bm{CZ}+\bm{CS}}$};
      \draw[thick,color2] (0.22cm,1.05cm) -- (0.28cm,1.2cm)
        node[inner sep=2pt,anchor=south west,fill=white] {\color{blue}$\bm{2\times \bm{CZ}+\bm{CT}}$};

      \draw[thick,color1] (-0.6cm,-0.85cm) -- (-0.7cm,-0.95cm)
        node[inner sep=2pt,anchor=north east,fill=white] {\color{red}$\bm{2\times \bm{CZ}+\bm{CS}}$};
      \draw[thick,color2] (-0.22cm,-1.05cm) -- (-0.28cm,-1.2cm)
        node[inner sep=2pt,anchor=north east,fill=white] {\color{blue}$\bm{2\times \bm{CZ}+\bm{CT}}$};

      \draw[<-] (-0.05cm,1.05cm) -- (-0.15cm,1.15cm)
        node[inner sep=1pt,anchor=south east,fill=white] {$\fswap_{\pi/2}=\iswap: 2\times\cz/$};
      \draw[<-] (0.05cm,-1.05cm) -- (0.15cm,-1.15cm)
        node[inner sep=1pt,anchor=north west,fill=white] {$\fswap_{-\pi/2}={\iswap}^\dagger: 2\times\cz/$};
      \draw[<-] (1.05cm,-0.05cm) -- (1.15cm,-0.15cm)
        node[inner sep=1pt,anchor=north west,fill=white] {$\fswap_0=\swap$};
      \draw[<-] (-1.05cm,-0.05cm) -- (-1.15cm,-0.15cm)
        node[inner sep=1pt,anchor=north east,fill=white] {$\fswap_\pi\sim\swap$};

      \draw[ultra thick,line width=4.5pt,color2] ( 67.5:1cm) arc ( 67.5: 112.5:1cm);
      \draw[ultra thick,line width=4.5pt,color2] (-67.5:1cm) arc (-67.5:-112.5:1cm);
      \draw[ultra thick,dashed,color1] ( 45:1cm) arc ( 45: 135:1cm);
      \draw[ultra thick,dashed,color1] (-45:1cm) arc (-45:-135:1cm);
      
      \filldraw[black] (  0:1cm) circle(1.2pt);
      \filldraw[black] ( 90:1cm) circle(1.2pt);
      \filldraw[black] (180:1cm) circle(1.2pt);
      \filldraw[black] (270:1cm) circle(1.2pt);

      \draw[line width=2.5pt,color2] (  67.5:0.95cm) -- (  67.5:1.05cm);
      \draw[line width=2.5pt,color2] ( 112.5:0.95cm) -- ( 112.5:1.05cm);
      \draw[line width=2.5pt,color2] ( -67.5:0.95cm) -- ( -67.5:1.05cm);
      \draw[line width=2.5pt,color2] (-112.5:0.95cm) -- (-112.5:1.05cm);

      \draw[ultra thick,color1]  (45:0.95cm) --  (45:1.05cm);
      \draw[ultra thick,color1] (135:0.95cm) -- (135:1.05cm);
      \draw[ultra thick,color1] (225:0.95cm) -- (225:1.05cm);
      \draw[ultra thick,color1] (315:0.95cm) -- (315:1.05cm);

      \draw (360:0.70cm) node[fill=white] {$\theta=0$};
      \draw (180:0.80cm) node[fill=white] {$\pi$};
      \draw (270:0.80cm) node[fill=white] {$-\pi/2$};
      \draw  (90:0.80cm) node[fill=white] {$\pi/2$};

      \draw  (45:0.80cm) node {$\pi/4$};
      \draw (135:0.80cm) node {$3\pi/4$};
      \draw (225:0.80cm) node {$-3\pi/4$};
      \draw (314:0.80cm) node {$-\pi/4$};
    \end{tikzpicture}
  \caption{\cz/-depth of optimized $\fswap_\theta$ decompositions. The angle $\theta$ defines the fermionic SWAP unitary, with $\theta = 0$ ($\pi$) corresponding to the standard SWAP gate (up to virtual phases), and $\theta = \pi/2$ ($-\pi/2$) corresponding to the $\iswap$ ($\iswap^\dagger$) gate. All $\fswap_\theta$ can be implemented with three \cz/ gates using Eq.~\ref{eq:fswap:cz}; however, only two \cz/ gates are needed for $\iswap$ and $\iswap^\dagger$ gates. For $\theta \in [\pm\pi/4, \; \pm3\pi/4]$ (red), $\fswap_\theta$ can be implemented with two $CZ$ and one $CS$ gate; whereas for $\theta \in [\pm3\pi/8, \; \pm5\pi/8]$ (blue), $\fswap_\theta$ can be implemented with two $CZ$ and one $CT$ gate.} 
  \label{fig:fswap:unit-circle}
  \end{center}
  \vspace{-5pt}
\end{figure}
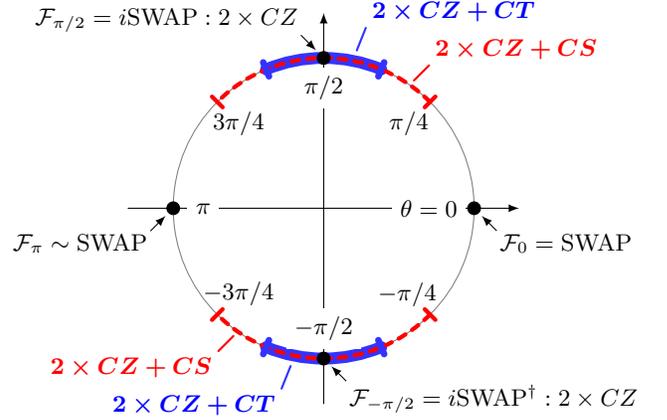

\Cref{eq:fswap:crz,eq:fswap:cz} suggest that a $\cphase{\phi}$ gate is also sufficient to generate any $\fswap_\theta$ for $|2\theta-\pi|\le\phi$. As shown in Fig.~\ref{fig:fswap:unit-circle}, by choosing $\phi=\pi/2$ (the \cs/ gate) we can implement half of all possible $\fswap_\theta$ gates with the equivalent of 2.5 \cz/ gates (that is, two \cz/ gates and one \cs/ gate).
For any $\pi/4\le\theta\bmod\pi\le3\pi/4$, we have,
\begin{equation}
\label{eq:fswap:cs}
{\Qcircuit @C=.5em @R=.2em {
  & \gate{X_{\mu}} & \ctrl{1} & \gate{\xx} & \ctrl{1} & \qw            & \gate{\xx} & \ctrl{1} 
  & \gate{Z_{5\pi/4}} & \qw & \qw \\
  & \gate{\xx}        & \ctrl{0} & \gate{\xx} & \ctrl{0} & \gate{Z_\lambda} & \gate{\xx} & \gate{S} 
  & \gate{Z_{5\pi/4}} & \gate{X_\nu} & \qw \\
}}_,
\end{equation}
where,
\begin{align}
\mu &= \csc^{-1}\big(-\sqrt2\sin\theta\big), \\
\lambda  &= \cos^{-1}\big(-\sqrt2\cos\theta\big), \\
\nu &= - 2\cos^{-1}\big(\sqrt{\cot\theta+1}/\sqrt2\big).
\end{align}
An equivalent decomposition can easily be found using a \csd/ gate in place of \cs/. The $X_{\mu}$ and $X_\nu$ gates can each be implemented with two $\xx$ pulses and virtual phases using Eq.~\ref{eq:zxzxz}; however, this complexity can be mitigated within a full fermionic SWAP network (\cref{sec:fswap:schedule}).

Additional controlled-$Z_\phi$ operations for fixed values of $\phi$ would further refine the optimized $\fswap_\theta$ decomposition toward the lower bound provided by fully-parameterized $\cphase{\phi}$. For example, as shown in Fig.~\ref{fig:fswap:unit-circle} one quarter of all possible $\fswap_\theta$ are reachable using a $\ct/=\cphase{\pi/4}=\sqrt[4]{\cz/}$ gate, and so the addition of a \ct/ to the gateset would reduce the \cz/ depth of these decompositions to the equivalent of 2.25 \cz/ gates.

\subsection{Optimized Fermionic SWAP Networks}
\label{sec:fswap:schedule}

We can further simplify the decomposition of fermionic SWAP gates in the context of the larger network. Various discrete and continuous symmetries in the fermionic SWAP operation result in degrees of freedom to its optimized decomposition:
\begin{enumerate}
    \item $\fswap_\theta = (X\otimes X)\fswap_\theta(X\otimes X)$,
    \item $\fswap_\theta = (1\otimes X)\fswap_{-\theta}(X\otimes 1)$,
    \item $\fswap_\theta = (Z\otimes Z)\fswap_{\theta+\pi}$,
    \item $\fswap_\theta(q_0,q_1) = \fswap_\theta(q_1,q_0)$ (qubit interchange),
    \item $\fswap_\theta = {\fswap}^\dagger_{-\theta}$,
    \item 
    $\fswap_\theta = (Z_\vartheta\otimes Z_\varphi) \fswap_\theta(Z_{-\varphi}\otimes Z_{-\vartheta})\;\forall\;\vartheta,\varphi\in\mathbb{R}$,
\end{enumerate}
where $\vartheta$, $\varphi$ are continuous parameters.
Symmetry 5 is useful only for $\fswap_\theta$ gates implemented using a  \cs/ or \csd/, in which case it can be used to reverse the order of entangling gates (and corresponding single-qubit gates $X_\mu$ and $X_\nu$) in the circuit.
(For the 3-\cz/ decomposition of $\fswap_\theta$ the physical implementations of $\fswap_\theta$ and $\fswap^\dagger_{-\theta}$ are identical.)
When implementing a fermionic SWAP network, these degrees of freedom can be exploited to maximize cancellation of single-qubit gates between sequential fermionic SWAP gates.

We use an automated scheduler which computes the set of logically equivalent decompositions (generated from symmetries 1-5) of each gate in the fermionic SWAP network, and searches for the sequence of decompositions which minimizes circuit depth (corresponding to $\xx$ count in the critical path). The continuous parameters $\vartheta,\varphi$ (symmetry 6) are numerically optimized for each gate in the sequence. Though in general this search space grows exponentially, in practice for the circuits used in this paper a simple best-first search quickly converges on an optimal or near-optimal circuit.

\section{Cycle Benchmarking of Optimized Decompositions}\label{sec:cycle_benchmarking}
\begin{figure}[h]
\centering
\includegraphics[width=\columnwidth]{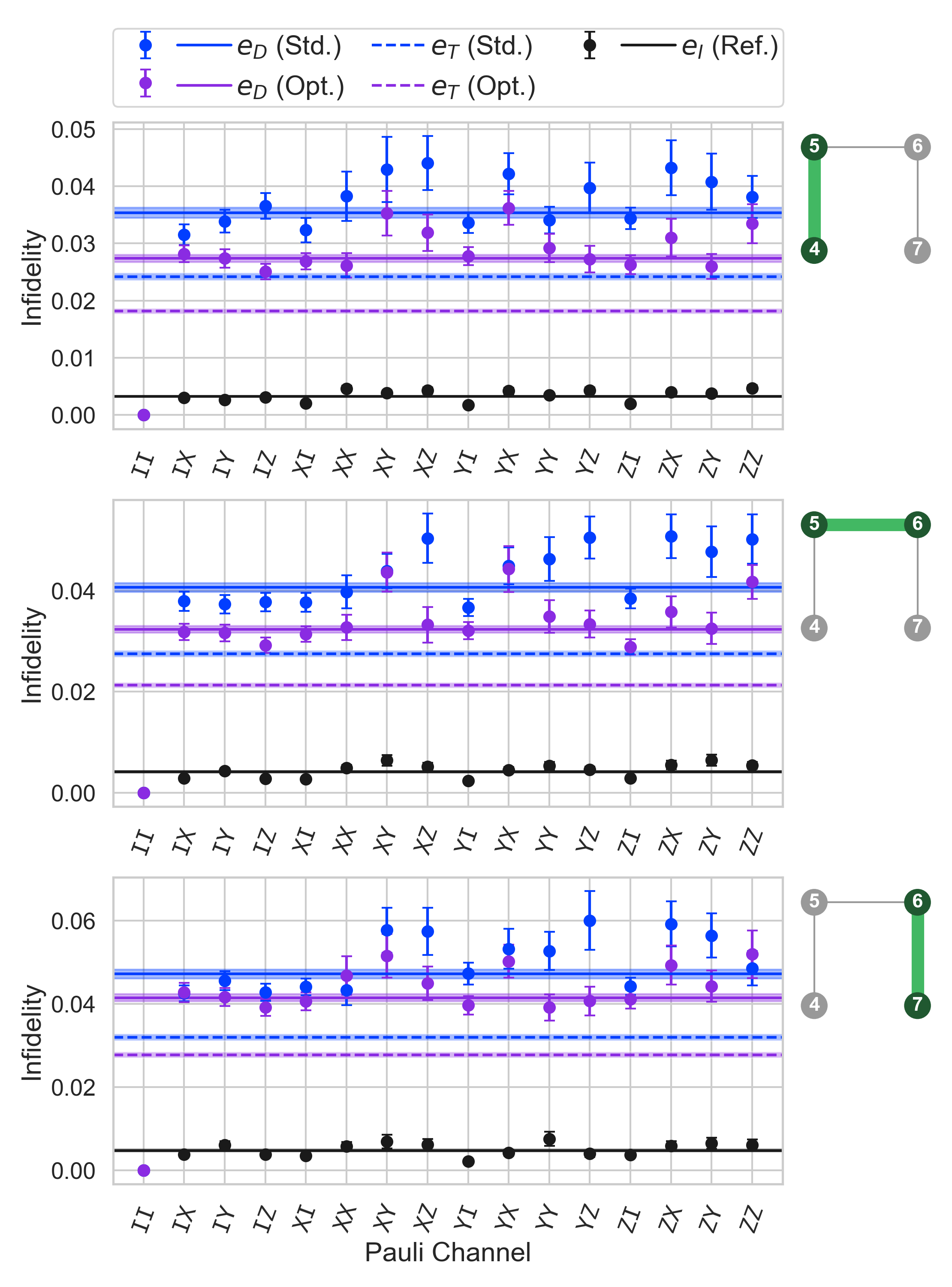}
\caption{Improved SWAP gates via gate-based optimizations. The process infidelity of the optimized SWAP gate between Q4 \& Q5 (top), Q5 \& Q6 (middle), and Q6 \& Q7 (bottom) is lower than the process infidelity of the standard decomposition due to the elimination of unnecessary gates. The blue, purple, and black data points represent the Pauli infidelity $1 -p_P$ for each Pauli channel $P$ for the standard (Std.) SWAP gate, optimized (Opt.) SWAP gate, and reference (Ref.) cycle, respectively. The solid blue (purple) line is the average process infidelity $e_D$ of the dressed cycle for the standard (optimized) SWAP gate, and the solid black line is the average process infidelity $e_I$ of the Pauli twirling operators. The dashed blue (purple) line is the process infidelity $e_T$ of target cycle (i.e.~SWAP gate) estimated via Eq.~\ref{eq:cb_eT} for the standard (optimized) SWAP sequence. The semi-transparent bands around the average process infidelities represents the 95\% confidence interval uncertainty of the estimates.}
\label{fig:cb_swap}
\end{figure}

To benchmark the performance of the optimized pulse sequences relative to their standard decompositions, we utilize cycle benchmarking \cite{erhard2019characterizing} (CB), a scalable protocol for measuring the performance of parallel gate cycles. Cycle benchmarking differs from randomized benchmarking \cite{emerson2005scalable, knill2008randomized, dankert2009exact, magesan2011scalable} (RB) in two keys ways: (i) it utilizes Pauli twirling instead of Clifford twirling, which maps gate errors into a stochastic Pauli channel (instead of a global depolarizing channel); (ii) CB benchmarks the performance of quantum gates performed in parallel, providing a measure of their performance in the context of multi-qubit quantum algorithms. In contrast, benchmarking the individual constituent gates of multi-qubit cycles has been shown to be a poor predictor of the global performance of quantum circuits \cite{proctor2020measuring} due to the presence of coherent errors and crosstalk between qubits, and because such benchmarks fail to capture errors on (or incurred by) idling spectator qubits \cite{krinner2020benchmarking}.

CB measures the process fidelity of a target cycle by preparing the system in a Pauli basis state (e.g. $XYIZ$ for four qubits), and measuring the exponential decay as a function of sequence depth. A separate exponential decay of the form $Ap_P$ can be fit for each basis preparation and measurement state $P$ (i.e. Pauli channel), where $A$ is the state-preparation and measurement (SPAM) parameter, and $p$ the fit parameter (i.e. process fidelity). Much like interleaved randomized benchmarking \cite{magesan2012efficient} (IRB), in which the target gate is interleaved between random Clifford gates, CB interleaves the target cycle between cycles of random single-qubit Pauli gates. Therefore, CB measures the process fidelity of a dressed cycle, which contains the errors due to the interleaved target cycle as well as the Pauli twirling gates. The total process fidelity is the average over $K$ Pauli channels,
\begin{equation}
    F = \frac{1}{K}\sum_{P \in \mathcal{P}} p_P,
\end{equation}
where the number of Pauli channels $K = |\mathcal{P}| \le 4^n$ ($n$ qubits) in the set $\mathcal{P}$ that are sampled out of the full $4^n$ possible states sets the precision of the fidelity estimate \cite{erhard2019characterizing}. The process infidelity of the dressed cycle is therefore given as $e_D = 1 - F$. To separate the infidelity of the target cycle from the twirling gates, we measure the CB fidelity of the ``all-identity" reference cycle, which equates to benchmarking the average performance of only the Pauli twirling gates. Similar to IRB, we can use this to estimate the process infidelity of the target ($T$) cycle by taking the ratio of the process fidelities of the dressed ($D$) and reference ($I$) cycles,
\begin{equation}\label{eq:cb_eT}
    e_T = \frac{d - 1}{d} \bigg(1 - \frac{F_D}{F_I} \bigg),
\end{equation}
where $d=2^n$ ($n$ qubits) is the dimension of the system. Using CB has been shown to tighten the upper- and lower-bounds on the fidelity estimate of the interleaved cycle relative to IRB \cite{mitchell2021hardware}, which can span orders of magnitude \cite{carignan2019bounding}. We use this method to estimate a target infidelities for $CZ$, $CS$, and $CS^\dagger$ gates (see Table \ref{tab:table_tqp} in Appendix \ref{appendix_b}).

\begin{table*}
  \centering
  \begin{tabular}{|c||c||c|c||c|c||c|c||c|c||c|c|}
        \hline
        \begin{tabular}{c}
        Cycle
        \end{tabular} &
        \Qcircuit @C=.5em @R=.15em {
            & \gate{I} & \qw \\
            & \gate{I} & \qw \\
            & \gate{I} & \qw \\
            & \gate{I} & \qw \\ \\
        } &
        \Qcircuit @C=.5em @R=.15em {
            & \gate{H} & \qw \\
            & \gate{H} & \qw \\
            & \gate{H} & \qw \\
            & \gate{H} & \qw \\ \\
        } &
        \Qcircuit @C=.5em @R=.15em {
            & \gate{H} & \qw \\
            & \gate{H} & \qw \\
            & \gate{H} & \qw \\
            & \gate{H} & \qw \\ \\
        } &
        \Qcircuit @C=.5em @R=.84em {
            & \gate{I} & \qw \\
            & \ctrl{1} & \qw \\
            & \ctrl{-1} & \qw \\
            & \gate{I} & \qw \\
        } & 
        \Qcircuit @C=.5em @R=0.49em {
            & \gate{I} & \qw \\
            & \gate{S} & \qw \\
            & \ctrl{-1} & \qw \\
            & \gate{I} & \qw \\
        } &
        \Qcircuit @C=.5em @R=1.2em {
            & \ctrl{1} & \qw \\
            & \ctrl{-1} & \qw \\
            & \ctrl{1} & \qw \\
            & \ctrl{-1} & \qw \\
        } & 
        \Qcircuit @C=.5em @R=.65em {
            & \gate{S^\dagger} & \qw \\
            & \ctrl{-1} & \qw \\
            & \gate{S^\dagger} & \qw \\
            & \ctrl{-1} & \qw \\
        } & 
        \Qcircuit @C=.5em @R=1.0em {
            & \gate{I} & \qw \\
            & \qswap\qwx[1] & \qw \\
            & \qswap\qwx[-1] & \qw \\
            & \gate{I} & \qw \\
        } & 
        \Qcircuit @C=.5em @R=1.0em {
            & \gate{I} & \qw \\
            & \qswap\qwx[1] & \qw \\
            & \qswap\qwx[-1] & \qw \\
            & \gate{I} & \qw \\
        } &
        \Qcircuit @C=.5em @R=1.4em {
            & \qswap\qwx[1] & \qw \\
            & \qswap\qwx[-1] & \qw \\
            & \qswap\qwx[1] & \qw \\
            & \qswap\qwx[-1] & \qw \\
        } & 
        \Qcircuit @C=.5em @R=1.4em {
            & \qswap\qwx[1] & \qw \\
            & \qswap\qwx[-1] & \qw \\
            & \qswap\qwx[1] & \qw \\
            & \qswap\qwx[-1] & \qw \\
        } \\
    \hline
    \hline
    Error Rate & Ref. & Std. & Opt. & Std. & Opt. & Std. & Opt. & Std. & Opt. & Std. & Opt. \\
    \hline
    $e_I$ (\num{e-3}) & 9.6(6) & & & & & & & & & & \\
    $e_D$ (\num{e-2}) & & 1.5(1) & 1.16(6) & 2.11(7) & 1.67(8) & 3.4(1) & 2.09(8) & 9.6(7) & 6.3(2) & 11.7(4) & 10.4(4) \\
    $e_T$ (\num{e-2}) & & 0.5(1) & 0.19(8) & 1.09(9) & 0.68(9) & 2.3(1) & 1.07(9) & 8.1(7) & 5.1(2) & 10.2(3) & 9.0(4) \\
    \hline
    \hline
    \multicolumn{2}{|c||}{Reduction in $e_T$} & \multicolumn{2}{c||}{64\%} & \multicolumn{2}{c||}{38\%} & \multicolumn{2}{c||}{53\%} & \multicolumn{2}{c||}{38\%} & \multicolumn{2}{c|}{12\%} \\
    \hline
  \end{tabular}
\caption{Benchmarked improvements in optimized cycles. All optimized (Opt.) cycles have a lower CB process infidelity than their respective standard (Std.) decompositions.}\label{tab:cb_eF}
\end{table*}

In Fig.~\ref{fig:cb_swap}, we plot the CB results for the standard and optimized SWAP gates (see Eq.~\ref{eq:circuit:swap}) between all three qubit pairs. We see that optimized target cycle infidelity $e_T$ of the SWAP gates is reduced 25\%, 23\%, and 13\% relative to the standard SWAP gate for (Q4, Q5), (Q5, Q6), and (Q6, Q7), respectively. This average improvement can generally be expected for circuits utilizing basic SWAP gates; however, further optimizations can be implemented in the context of full fermionic SWAP networks with the replacement of one of the $CZ$s in the SWAP gate with a $CS$ or $CS^\dagger$ gate, as outlined in the previous section. Furthermore, while the benchmarking results in Fig.~\ref{fig:cb_swap} show improvements in the SWAP gates between all qubit pairs, they do not capture what improvements can be expected for cycles of gates in any four-qubit application. In Table \ref{tab:cb_eF}, we compare the benchmarked process infidelities of optimized gate cycles versus the standard decompositions for relevant cycles appearing in four-qubit fermionic SWAP networks (see \cref{sec:qaoa}). These include the all-Hadamard cycle for basis preparation and converting $CZ$s to $CX$s, the relevant multi-qubit gate cycles appearing in the fermionic SWAP networks, and the parallel SWAP cycles incorporating the optimizations outlined in the previous section. We see universal improvement in the target cycle infidelity $e_T$ for the optimized cycles, with reductions in $e_T$ ranging from 64\% for the all-Hadamard cycle to 12\% for the SWAP cycle. These results demonstrate that simple improvements in circuit decomposition and gate optimizations can lead to dramatic improvements in benchmarked gate and cycle performance. Next, we highlight how these fidelity improvements can lead to performance improvements in fermionic SWAP networks.

\section{Application Benchmarking of QAOA}\label{sec:qaoa}
The Quantum Approximate Optimization Algorithm (QAOA) \cite{farhi2014quantum} describes a variational ansatz for solving combinatorial optimization problems described by an objective Hamiltonian $H$. QAOA is characterized by a hyperparameter $p$ that specifies the depth of the ansatz. Specifically, the ansatz is $e^{i\beta_p B}e^{i\gamma_p H}...e^{i\beta_1 B}e^{i\gamma_1 H}$, where $B=\sum_{i}X_i$ is a mixing Hamiltonian and $\vec\gamma$, $\vec\beta$ represent $2p$ classically optimized variational parameters. It is believed that QAOA is hard to approximate even at $p=1$ and is therefore a leading candidate for demonstrations of quantum advantage \cite{farhi2016quantum}. We generate QAOA circuits corresponding to Sherrington-Kirkpatrick spin-glass model Hamiltonians with edge weights $J_{ij}$ randomly selected from $\pm1$ (see \cref{appendix_c} for the exact symbolic form of the circuits). Each $e^{i\gamma H}$ is then implemented with fermionic SWAP networks with gates $\fswap_{\pm\gamma}$. Parameters $\beta_i,\gamma_i$ are sampled uniformly from $[0,2\pi)$.

In Fig.~\ref{fig:fswap_combined}, we measure two-qubit ($p=1$) and four-qubit ($p=1$ and $p=2$) QAOA circuits (see \cref{appendix_c} for example circuits) for various angles $\gamma$ and benchmark the performance using the total variation distance (TVD),
\begin{equation}\label{tvd}
    D(p, q) = \frac{1}{2}\sum_{x \in X} \vert p_x - q_x \vert,
\end{equation}
where $p_x$ is the probability of measuring a bit string $x$ in a set $X$, and $q_x$ is the ideal (noiseless) probability. We see that the optimized (Opt.) circuits generally provide more accurate performance relative to the standard (Std.) decompositions, reducing the average TVD from $D_\text{Std.} = 0.20(5)$ to $D_\text{Opt.} = 0.14(3)$ for four-qubit QAOA circuits of depth $p=1$, and from $D_\text{Std.} = 0.23(4)$ to $D_\text{Opt.} = 0.22(6)$ for circuits of depth $p=2$. For two-qubit networks, the optimized circuits outperform the standard circuits on average for qubits (Q5, Q6), but perform worse [equivalent] for (Q4, Q5) [(Q6, Q7)]. We conjecture that the failure of the optimized circuits to outperform the standard circuits for qubits (Q4, Q5) and (Q6, Q7) is due to systemic coherent errors, whose impacts can dominate algorithm performance and are not accurately captured by randomized benchmarks (see the discussion in Section \ref{sec:equivalent_circuit_averaging}). The parameter angle $\gamma$ determines what gate optimizations can be implemented for each network, with $\pi/4 \leq \gamma \leq 3\pi/4$ and $5\pi/4 \leq \gamma \leq 7\pi/4$ defining the angles for which $CS$ or $CS^\dagger$ gates can be used in place of $CZ$ gates (see Fig.~\ref{fig:fswap:unit-circle}). The $\gamma$ values are randomly chosen in Fig.~\ref{fig:fswap_combined}, but they are seeded such that half of the fermionic SWAP networks tested can take advantage of the $CS$ or $CS^\dagger$ gates at $p=1$. These results demonstrate that simple changes to circuit decomposition and gate-based optimizations can lead to clear improvements in algorithm and application performance, highlighting the importance of smart compilers and more continuous gatesets in the NISQ era.

\begin{figure}[!ht] 
    \centering
    \includegraphics[width=\columnwidth]{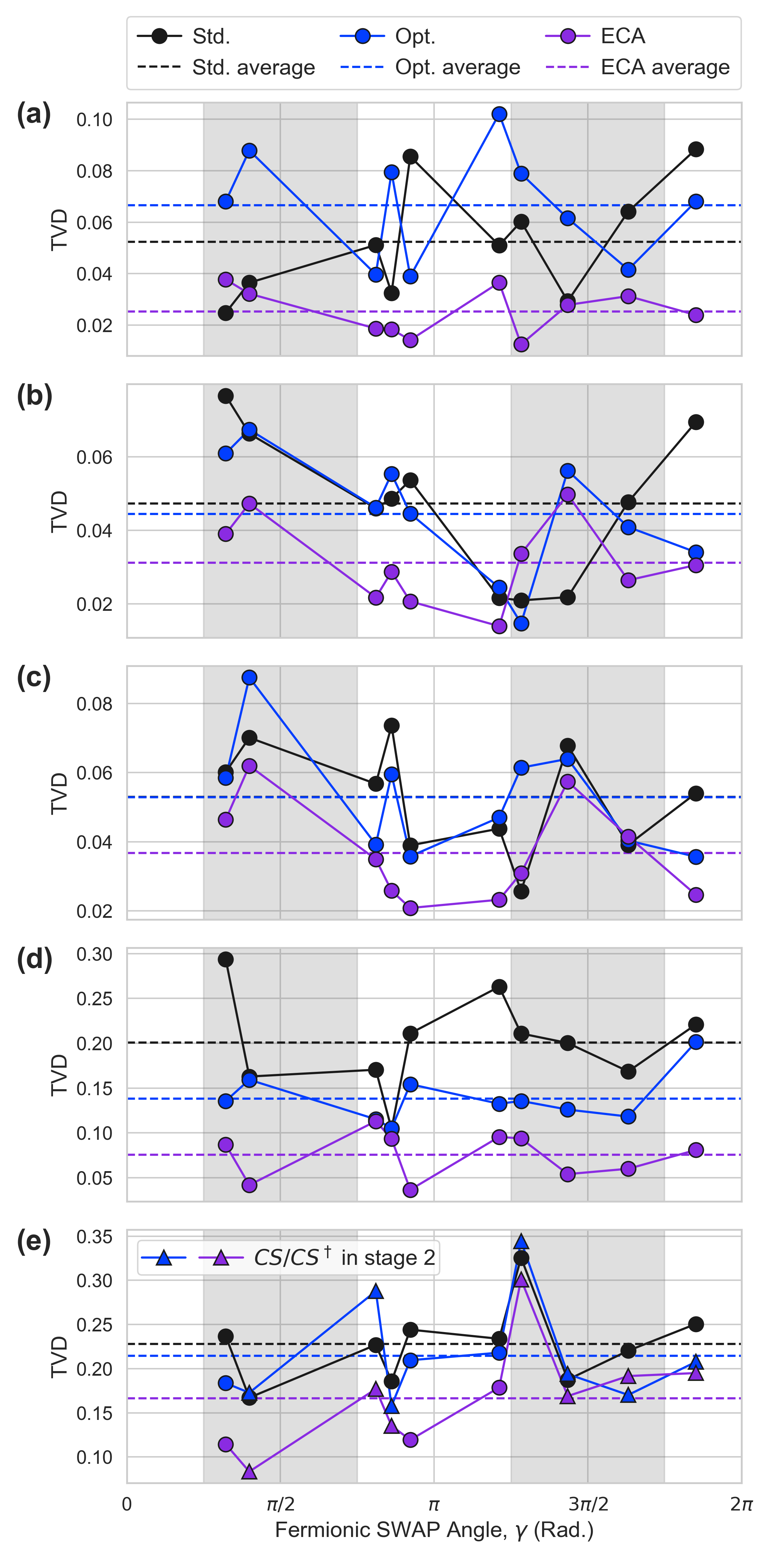}
    \caption{Improved fermionic SWAP networks via gate-based optimizations. The TVD performance of fermionic SWAP networks of angle $\gamma$ are plotted for qubits (a) (Q4, Q5), (b) (Q5, Q6), and (c) (Q6, Q7), and four-qubit circuits with (d) $p=1$ and (e) $p=2$ stages. For (b), (d), and (e), the optimized (blue, Opt.) circuits outperform the standard (black, Std.) on average (dashed lines). The ECA (purple) results consistently outperform both the standard and optimized circuits. The grey shaded regions define the angles for which $CS$ or $CS^\dagger$ gates can be utilized. The results in (d) are plotted against the $\gamma$ from stage 1 (the triangle markers denote circuits which utilize a $CS$ or $CS^\dagger$ gate in stage 2). (Error bars on the TVD $\sim \mathcal{O}[10^{-3}]$ are smaller than the markers.)}
\label{fig:fswap_combined}
\end{figure}

\section{Equivalent Circuit Averaging}\label{sec:equivalent_circuit_averaging}
One limitation of benchmarking the average performance of gates or cycles is that randomized benchmarks are not accurate predictors of the global performance of structured quantum circuits due to the presence of coherent errors \cite{proctor2020measuring}. When averaging over a twirling group, such as the Clifford (Pauli) group for RB (CB), all errors are converted into a global depolarizing (stochastic Pauli) channel. However, in actual quantum algorithms, the physical error mechanisms are more complex than depolarizing or Pauli channels, as coherent errors can interfere constructively or destructively from one cycle to the next. Therefore, while the optimized pulse sequences show clear improvements in cycle fidelity (cf.~Table~\ref{tab:cb_eF}) measured via CB, this does not always guarantee improvements in algorithm performance composed of these cycles. This can be seen in Fig.~\ref{fig:fswap_combined}, in which the standard circuit decompositions occasionally outperform the optimized circuits for the four-qubit results, and outperform the optimized circuits on average for the two-qubit results for qubits (Q4, Q5).

Being systematic in nature, coherent errors can in theory be measured and corrected via recalibration or added compensation pulses. However, the complexity of fully characterizing coherent errors (i.e.~context-dependent rotation axes and angles \cite{rudinger2021experimental}) on multiqubit processors that arise due to classical and quantum crosstalk is intractable, and no known scalable methods exist for doing so for systems with continuous single-qubit gatesets. Various methods exist for suppressing coherent errors, such as dynamical decoupling \cite{tripathi2021suppression} and error-correcting composite pulse sequences \cite{low2014optimal}, or randomization methods for ``tailoring" them into stochastic noise, such as Pauli twirling \cite{geller2013efficient, cai2020mitigating, song2019quantum, kim2021scalable}, Pauli frame randomization \cite{knill2004fault, kern2005quantum, ware2021experimental}, and randomized compiling \cite{wallman2016noise, hashim2020randomized}. However, these methods generally require the modification of single-qubit gates or the inclusion of more gates (e.g. in the case of dynamical decoupling and composite sequences), or require that the two-qubit gates in circuits are Clifford so that inverting Pauli operators can be efficiently computed and applied. Adopting these techniques would therefore require forgoing the circuit optimizations (and corresponding fidelity gains) employed so far in this work --- both by necessitating additional $\xx$ pulses and precluding the use of non-Clifford $CS$ and $CS^\dagger$ gates.

A similar strategy has been proposed for circuit synthesis methods, in which systematic approximation errors are rendered incoherent by averaging over various circuits near a target unitary generated from ensembles of approximate decompositions \cite{hastings2016turning, campbell2017shorter}. We employ this general idea (with systematic errors in the physical gates taking the place of approximation errors) using the space of equivalent fermionic SWAP decompositions generated by the degrees of freedom outlined in \cref{sec:fswap:schedule}. If we were to randomly sample from these decompositions for each gate in a circuit, we would forgo the single-qubit gate reduction achieved by the optimized scheduling procedure. However, the sequence of gate decompositions which minimizes the depth of the overall circuit is in general not unique. By treating the circuit holistically, we can sample from the subset of logically equivalent circuits which all minimize circuit depth. This is easily implemented by randomizing the search path used by the scheduler. Though the constraints on randomization necessary to preserve circuit depth mean that we cannot make solid guarantees on the mitigation of coherent errors, we empirically find that averaging over equivalent circuits generated in this way is an effective strategy for systematic error mitigation. We call this strategy Equivalent Circuit Averaging (ECA).

For the circuits in Fig.~\ref{fig:fswap_combined}, we generate $M=20$ logically equivalent optimized circuits for each angle $\gamma$ (see \cref{appendix_c} for example circuits). In order to normalize shot statistics, we measure each equivalent circuit $s = S/M$ times and compute the union over all $M$ results to obtain an equivalent statistical distribution for a circuit measured $S$ times; $S = 10000$ and $s = 500$ for the results in Fig.~\ref{fig:fswap_combined}. We see that ECA dramatically reduces the TVD on average in comparison to both the standard and optimized results for all of the two- and four-qubit fermionic SWAP network results, reducing the average TVD by $\sim 60\%$ [$26\%$] from $D_\text{Std.} = 0.20(5)$ to $D_\text{ECA} = 0.08(2)$ [$D_\text{Std.} = 0.23(4)$ to $D_\text{ECA} = 0.17(6)$] for the four-qubit $p=1$ [$p=2$] QAOA results, and providing the most accurate measured probability distribution in 88\% of all of the two- and four-qubit circuits measured.

While the classical overhead of generating and measuring $M$ logically equivalent circuits scales roughly linearly in $M$, we observe significant improvements in the measured results. These results demonstrate that ECA is a useful tool for smart compilers which optimize circuit decomposition using various degrees of freedom, and is not limited to circuits only containing two-qubit Clifford gates, adding to the toolbox of randomization methods that can be employed in the NISQ era.

\section{Conclusions} \label{sec:conclusion}
Quantum compilers play a fundamental role in the translation of abstract quantum circuits to machine instructions in gate-based quantum computing. In the NISQ era, it is necessary to consider the balance between the calibration overhead for large gatesets and optimal circuit decomposition for quantum application performance. In this work, we show that utilizing a smart compiler for cancelling unnecessary single-qubit gates is a simple method for lowering gate error rates in quantum circuits. We further demonstrate that by adding an additional two-qubit gate ($CS$ or $CS^\dagger$) to our gateset for each qubit pair, we observe significant improvement in benchmarked cycle and application performance. While our work focuses on fermionic SWAP networks and their application to QAOA, non-Clifford $CS$ gates also find importance in universal quantum computation and magic-state distillation for fault-tolerance \cite{cross2016scalable, haah2018codes, glaudell2021optimal}.

Additionally, we introduce Equivalent Circuit Averaging to mitigate the impact of systematic coherent errors in non-Clifford circuits by utilizing the various degrees of freedom of quantum compilers to generate many logically equivalent circuits. Given the difficulty in characterizing and predicting the impact of coherent errors on algorithm performance, such a method negates the need for doing so and assumes that the average over many circuits will reduce the impact of coherent errors on the algorithm results. We demonstrate the effectiveness of this approach with our application benchmark results, in which we find that ECA improves the accuracy of the measured probability distribution for 88\% of the randomly-generated two- and four-qubit QAOA circuits.

While ECA was employed by taking advantage of the various degrees of freedom of fermionic SWAP networks, a more sophisticated search procedure would likely expand the applicability of our methods for scheduling and generating equivalent circuits for more general applications. We further imagine possible ``hybrid'' strategies in which ECA is combined with other randomization protocols (e.g.~randomized compiling) for maximizing the ways in which logically equivalent circuits can be expressed, thus minimizing residual coherent errors. The cost of ECA (both classically and in terms of single-qubit gate optimization) in the general case and the degree to which it tailors noise in quantum systems (i.e.~in the manner of other randomization methods which twirl over a specific gateset) are open questions, which we plan to explore in future work.

Finally, as described in \cref{sec:optimized_fermionic_swap_gates}, access to a parameterized $\cphase\phi$ gate would minimize the \cz/ gate time for any fermionic SWAP gate. The corresponding gate decomposition (Eq.~\ref{eq:fswap:crz}) also avoids the $\theta$-dependent $X_\mu$ and $X_\nu$ gates in Eq.~\ref{eq:fswap:cs}, allowing for more efficient gate cancellation and a greater opportunity for randomness in ECA. The experimental validation of this decomposition would be a natural extension of this work, and would provide insight into the value of parameterized two-qubit gates for NISQ systems.

\section*{Acknowledgements}
This work was supported by the U.S. Department of Energy, Office of Science, Office of Advanced Scientific Computing Research Quantum Testbed Program under Contract No. DE-AC02-05CH11231. This material is also supported by the U.S. Department of Energy, Office of Science, Office of Advanced Scientific Computing Research under Award Number DE-SC0021526. A.H. acknowledges financial support from the National Defense Science \& Engineering Graduate (NDSEG) Fellowship.

R.R.~and P.G.~devised the optimized circuit decompositions. A.H.~conducted the experiments and analyzed the data. ECA was conceived by A.H.~and developed by R.R.~and P.G. R.K.N., D.I.S., F.C., and I.S.~supervised all theoretical and experimental work. J.M.K.~fabricated the sample. V.O. developed the SuperstaQ software interface to the AQT. A.H., R.R., V.O., and P.G.~wrote the manuscript with input from all coauthors.

F.C., P.G., R.R., and V.O.~have a financial interest in Super.tech and the SuperstaQ platform. F.C.~is also an advisor to Quantum Circuits, Inc. All other authors declare no competing interest.

\appendix
\setcounter{table}{0}
\renewcommand{\thetable}{A\arabic{table}}

\setcounter{figure}{0}
\renewcommand{\thefigure}{A\arabic{figure}}

\section{Single-qubit parameters}\label{appendix_a}
\begin{table}[h]
  \centering
  \resizebox{\columnwidth}{!}{
  \begin{tabular}{|l||r|r|r|r|}
    \hline
    {} & Q4 & Q5 & Q6 & Q7 \\
    \hline
    \hline
    Qubit freq.~(GHz) & 5.254275 & 5.331004 & 5.490952 & 5.661671 \\
    Anharm. (MHz) & -275 & -275 & -271.35 & -269 \\
    $T_1$ ($\mu$s) & 60(5) & 62(5) & 52(4) & 55(8) \\
    $T_2^*$ ($\mu$s) & 36(5) & 37(6) & 36(6) & 33(6) \\
    $T_2^\text{echo}$ ($\mu$s) & 62(5) & 73(7) & 68(7) & 54(6) \\
    Readout $P(0|0)$ & 0.999 & 0.995 & 0.995 & 0.995 \\
    Readout $P(1|1)$ & 0.990 & 0.989 & 0.979 & 0.974 \\
    RB (\num{e-3}) & 0.68(2) & 1.01(2) & 0.95(5) & 0.67(1) \\
    Sim. RB (\num{e-3}) & 1.49(8) & 2.5(2) & 3.1(2) & 2.4(2) \\
    \hline
  \end{tabular}}
\caption{Single-qubit parameters.}\label{tab:table_sqp}
\end{table}

Table \ref{tab:table_sqp} lists the relevant qubit parameters for the four transmon qubits used in this work. Qubit frequencies and anharmonicities are measured using Ramsey spectroscopy. Relaxation ($T_1$) and coherence ($T_2^*$ and $T_2^\text{echo}$) times are extracted by fitting exponential decay curves to the excited state lifetime and Ramsey spectroscopy measurements (without and with an echo pulse), respectively. Readout fidelities [$P(0|0)$ and $P(1|1)$] are determined by performing ensemble measurements of the qubits prepared in $\ket{0}$ and $\ket{1}$ and classifying the results using a Gaussian Mixture Model fit to the in-phase (I) and quadrature (Q) heterodyne voltage signals. Error rates for single-qubit gates are measured using randomized benchmarking (RB) and simultaneous (Sim.) RB. All error rates are defined in terms of the process infidelity $e_F = 1 - p$, where $p$ is the exponential fit parameter in $Ap^m$ for a sequence depth of $m$ and SPAM parameter $A$. This is equivalent to the average gate infidelity $r(\mathcal{E})$,
\begin{equation}\label{process_inf}
    e_F(\mathcal{E}) = r(\mathcal{E})\frac{d+1}{d},
\end{equation}
where $d=2^n$ is the system dimensionality ($n$ qubits).

\section{Two-qubit gate parameters}\label{appendix_b}
\begin{table}[h]
\centering
\resizebox{\columnwidth}{!}{
    \begin{tabular}{|l|l||r|r|r|}
        \hline
        \multicolumn{2}{|c||}{Gate \;\;\; / \;\;\; Qubits:} & (Q4, Q5) & (Q5, Q6) & (Q6, Q7) \\
        \hline
        \hline
        \multirow{5}{*}{$CZ$} & Duration (ns) & 200 & 200 & 200 \\
                            & Ramp fraction & 0.3 & 0.3 & 0.3 \\
                            & RB $e_F$ (\num{e-2}) & 1.9(1) & 2.04(8) & 1.95(6) \\
                            & CB $e_D$ (\num{e-2}) & 1.09(1) & 1.05(1) & 1.26(1) \\
                            & CB $e_T$ (\num{e-3}) & 5.8(1) & 4.8(1) & 5.9(2) \\
        \hline
        \multirow{4}{*}{$CS$} & Duration (ns) & & 150 & \\
                            & Ramp fraction &  & 0.4 & \\
                            & CB $e_D$ (\num{e-2}) & & 0.98(1) & \\
                            & CB $e_T$ (\num{e-3}) & & 4.3(1) & \\
        \hline
        \multirow{4}{*}{$CS^\dagger$} & Duration (ns) & 150 & & 150 \\
                            & Ramp fraction & 0.4 &  & 0.4  \\
                            & CB $e_D$ (\num{e-2}) & 0.98(1) & & 0.91(1) \\
                            & CB $e_T$ (\num{e-3}) & 5.0(1) & & 3.3(1) \\
        \hline
        \hline
        
        Ref. & CB $e_I$ (\num{e-3}) & 3.24(5) & 4.12(8) & 4.8(1) \\
        \hline
    \end{tabular}}
\caption{Two-qubit gate parameters.}\label{tab:table_tqp}
\end{table}
Table \ref{tab:table_tqp} lists the parameters for the individual $CZ$, $CS$, and $CS^\dagger$ gates used in this work. All two-qubit gates are composed of square pulses with cosine ramps. The total gate duration of each pulse (including the ramps) are listed in Table \ref{tab:table_tqp}; the fraction of the total gate duration for the ramp up and ramp down (individually) are specified under `Ramp fraction'. Although the $CS$ and $CS^\dagger$ gates can nominally be performed in half the duration of the $CZ$ gates, the cosine ramps limit the minimum duration of the gates. Instead, the $CS$ and $CS^\dagger$ gates are constructed to contain approximately half the total integrated area under the curve as the $CZ$ gates, thus performing half of the conditional rotation as the $CZ$. This is only approximate, since the conditional stark shift on each qubit will differ depending on the drive frequencies and amplitudes.

The choice of $CS$ versus $CS^\dagger$ for each qubit pair was determined depending on the sign of the Stark-induced ZZ interaction (cf.~Refs.~\cite{mitchell2021hardware, wei2021quantum}); it is more efficient (i.e., requires a smaller amplitude) to drive the $CS$ rotation in one direction for some qubits, and in the opposite direction for other qubits, depending on the detuning of the drive signal. While the $CZ$ can be benchmarked using RB, the $CS$ and $CS^\dagger$ gates are non-Clifford, and therefore require either non-Clifford RB \cite{cross2016scalable, garion2021experimental} or cycle benchmarking \cite{erhard2019characterizing} (CB) with refocusing pulses (used in this work). Table \ref{tab:table_tqp} lists the average process infidelity $e_D$ of the dressed cycle (target gate plus Pauli twirling gates), as well as the inferred process infidelity $e_T$ of the target gate alone (cf.~Eq.~\ref{eq:cb_eT}) using the measured CB process infidelity $e_I$ of the ``all-identity" reference cycle for each qubit pair. Two-qubit RB process infidelities $e_F$ are also included for the $CZ$ gates. While the fidelities of the individual two-qubit gates are useful for determining the quality of the gates in general, the process infidelities of the distinct parallel four-qubit cycles are more relevant to the application circuits presented in the body of this work. These values are listed in Table \ref{tab:cb_eF} of the main text.

\section{Example Fermionic SWAP network circuits}\label{appendix_c}

\begin{figure*}[h] 
    \centering
    \includegraphics[width=\textwidth]{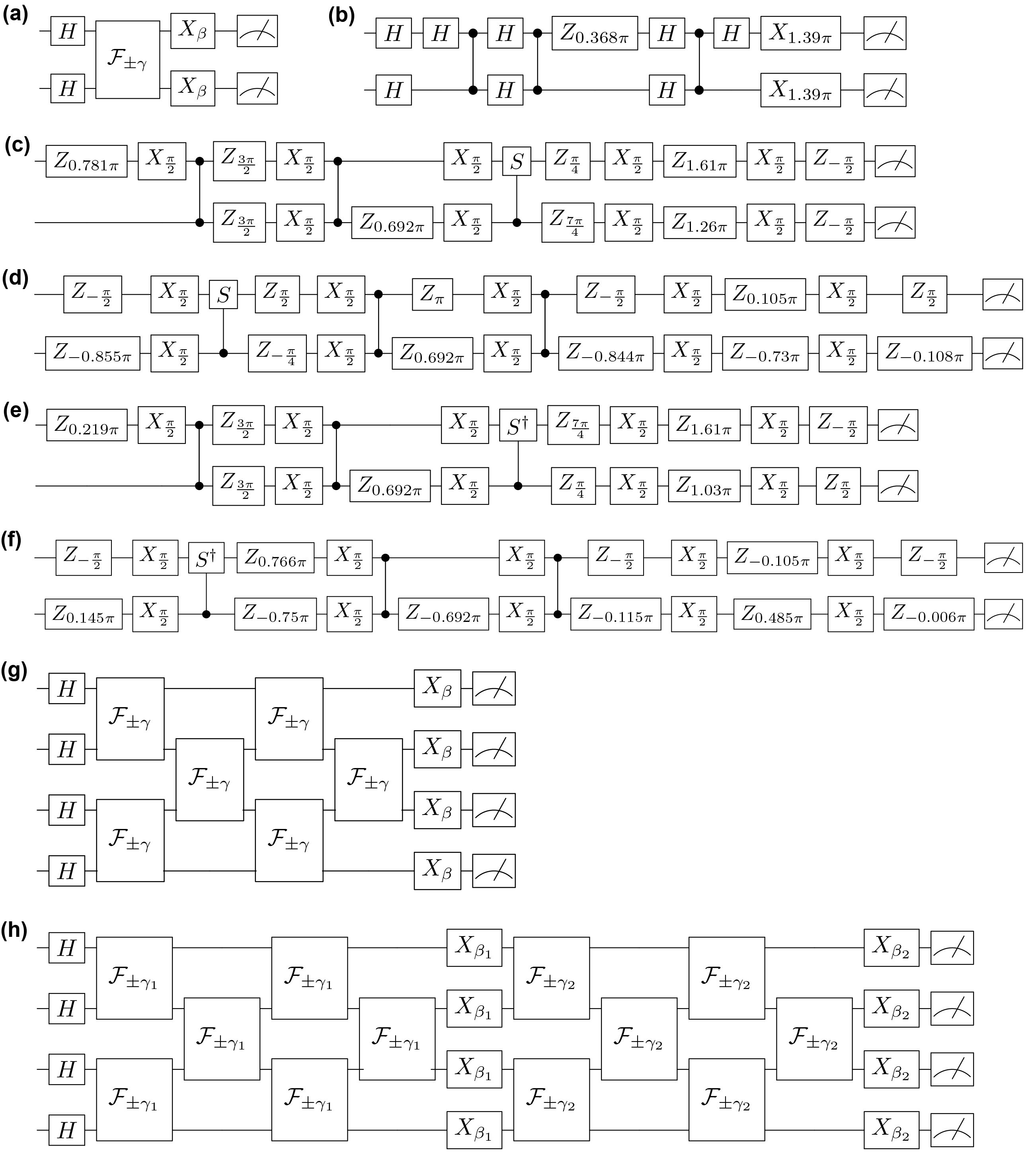}
    \caption{Example fermionic SWAP networks. Symbolic circuit representations of Fermionic SWAP networks (of depth $p$) for (a) two qubits ($p = 1)$, (g) four qubits ($p = 1)$, and (h) four qubits ($p = 2)$. (b) Baseline decomposition of a two-qubit fermionic SWAP network for a random choice of $\gamma$ and $\beta$ with $CZ$s. (c) Optimized decomposition of the circuit in (b) in terms of the native gateset utilizing a $CS$ instead of a $CZ$. (d) Logically equivalent decomposition of the circuit in (c). (e) Optimized decomposition of the circuit in (b) in terms of the native gateset utilizing a $CS^\dagger$ instead of a $CZ$. (f) Logically equivalent decomposition of the circuit in (e).}
\label{fig:eg_circuit_diagrams}
\end{figure*}

Example circuits for the fermionic SWAP networks presented in the main text can be seen in Fig.~\ref{fig:eg_circuit_diagrams}. This includes the symbolic representation of the two- and four-qubit Fermionic SWAP networks of depth $p=1$, and the four-qubit Fermionic SWAP network of depth $p=2$. An example two-qubit circuit is presented in Fig.~\ref{fig:eg_circuit_diagrams}(b) for a random choice of the two Fermionic SWAP angles, $\gamma$ and $\beta$. Additionally, the exact decompositions for this circuit in terms of $CS$ and $CS^\dagger$ gates are included, as well as logically equivalent variants of each.

\bibliography{refs}

\end{document}